\documentclass[prl,reprint,twocolumn,superscriptaddress,noshowpacs,notitlepage,longbibliography]{revtex4-1}%
\usepackage{graphicx,bm,times}
\usepackage{amsmath}
\usepackage{amsfonts}
\usepackage{amssymb}
\usepackage{color}
\usepackage{hyperref}
\hypersetup{
	colorlinks = true,
	allcolors = {blue}
}

\begin{document}

\title{Electronic anisotropies revealed by detwinned ARPES measurements of FeSe}

\author{Matthew D. Watson}
\email[corresponding author:]{mdw5@st-andrews.ac.uk}
\affiliation{Diamond Light Source, Harwell Campus, Didcot, OX11 0DE, United Kingdom}

\author{Amir A. Haghighirad}
\affiliation{Clarendon Laboratory, Department of Physics,
	University of Oxford, Parks Road, Oxford OX1 3PU, United Kingdom}

\author{Luke C. Rhodes}
\affiliation{Diamond Light Source, Harwell Campus, Didcot, OX11 0DE, United Kingdom}
\affiliation{Department of Physics, Royal Holloway, University of London, Egham, Surrey, TW20 0EX, United Kingdom}

\author{Moritz Hoesch}
\affiliation{Diamond Light Source, Harwell Campus, Didcot, OX11 0DE, United Kingdom}

\author{Timur K. Kim}
\email[corresponding author:]{timur.kim@diamond.ac.uk}
\affiliation{Diamond Light Source, Harwell Campus, Didcot, OX11 0DE, United Kingdom}

\begin{abstract}

 We report high resolution ARPES measurements of detwinned FeSe single crystals. The application of a mechanical strain is used to promote the volume fraction of one of the orthorhombic domains in the sample, which we estimate to be 80$\%$ detwinned. While the full structure of the electron pockets consisting of two crossed ellipses may be observed in the tetragonal phase at temperatures above 90~K, we find that remarkably, only one peanut-shaped electron pocket oriented along the longer $a$ axis contributes to the ARPES measurement at low temperatures in the nematic phase, with the expected pocket along $b$ being not observed. Thus the  low temperature Fermi surface of FeSe as experimentally determined by ARPES consists of one elliptical hole pocket and one orthogonally-oriented peanut-shaped electron pocket. Our measurements clarify the long-standing controversies over the interpretation of ARPES measurements of FeSe.   

\end{abstract}
\maketitle


\section{INTRODUCTION}

FeSe has emerged as a focus of research within the field of Fe-based superconductors due to the observations of high-$T_c$ in FeSe under pressure, in intercalates, and in monolayer form, and also because of the insight that it gives on the unresolved mystery of the nematic phase \cite{Coldea2017Review_arxiv}. FeSe undergoes a tetragonal-to-orthorhombic ``nematic" structural transition at 90~K, but unlike other parent compounds of Fe-based superconductors this is not accompanied by magnetic ordering at any temperature. While the symmetry-breaking distortion of the lattice is only weak \cite{Bohmer2014}, much more sizeable anisotropies are observed in measurements which probe electronic properties, such as  resistivity \cite{Tanatar2016}, polarized femtosecond pump–probe spectroscopy \cite{Luo2017} and quasiparticle interference \cite{Kasahara2014,Sprau2017Science}. It has been argued that orbital and not spin degrees of freedom are the driving force behind nematicity in FeSe \cite{Bohmer2014,Chubukov2016}, although perspectives that emphasize the role of magnetic fluctuations have also been suggested \cite{Glasbrenner2015,Mukherjee2015,Kreisel2015,Wang2016}. The desire to determine the magnitude and momentum-dependence of orbital ordering effects has motivated several angle-resolved photo-emission spectroscopy (ARPES) studies focusing on the evolution of the electronic structure through the nematic transition   \cite{Maletz2014,Nakayama2014,Watson2015a,Zhang2015,Fedorov2016,Watson2016,Fanfarillo2016,Watson2017c}. However, a significant challenge is that FeSe samples will naturally form structural twin domains in the nematic phase, leaving some ambiguity in the interpretation of the data because both domains contribute to the measured intensity and restore fourfold symmetry macroscopically. Only by measuring `detwinned' crystals \cite{Shimojima2014,Suzuki2015} can the underlying symmetry-breaking be fully revealed, and the controversies surrounding the interpretation of the data be put to rest.

 In this article we report high resolution ARPES measurements of single crystal samples of FeSe which are detwinned in the low-temperature orthorhombic phase by the application of a mechanical strain. We confirm that the longer axis of the elliptical hole pocket is parallel to the shorter $b$ axis of the orthorhombic structure. Our measurements of the electron pockets in one domain below $T_s$ reveal a stunning anisotropy: although all reasonable models of the electronic structure of FeSe to date have included electron pockets oriented along both the $a$ and $b$ directions, in the nematic phase only the peanut-shaped electron pocket oriented along the longer $a$ axis may be observed by ARPES. This remarkable symmetry-breaking in the observed electronic structure can be considered to be profound manifestation of nematic order.

\section{Methods}
Samples were grown by the chemical vapour transport method \cite{Bohmer2013,Watson2015a}. Samples of approximately 1000 $\times$ 1000 $\times$ 50 $\mu$m size with uniform thickness and regular facets were selected and mounted across the plates of the horseshoe-shaped device. The samples were aligned by eye (within $\sim 2^\circ$) such that the Fe-Fe direction matched the direction of strain. Epo-Tek H27D silver epoxy was used to mount the sample and also acts as a medium to transmit the strain into the sample. The devices were put under tension before sample mounting and then further strain was applied to samples once the epoxy is set, using the adjusting screw. From the pitch of the screw we estimate that the tensile strain is $\sim$ 1\% although the actual strain at the sample surface could be different due to imperfect coupling, differential thermal expansion and other effects. Due to the finite Poisson's ratio, the actual strain on the sample may be described as an anisotropic triaxial strain. The main text focuses on the sample where the greatest detwinning effect was observed, but information on further partially detwinned samples is provided in the Supplementary Information (SI). ARPES measurements were performed at the I05 beamline at the Diamond Light Source, UK. The photoelectron energy and angular distributions were analyzed with a SCIENTA R4000 hemispherical analyzer. The measurement temperature was 10 K unless explicitly stated, and the sample remained in a vacuum of $<2\times10^{-10}$ mBar throughout the measurements. The angular resolution was 0.2$^\circ$, and the overall energy resolution was better than 10 meV.

\section{RESULTS}

\begin{figure}
	\centering
	\includegraphics[width=\linewidth]{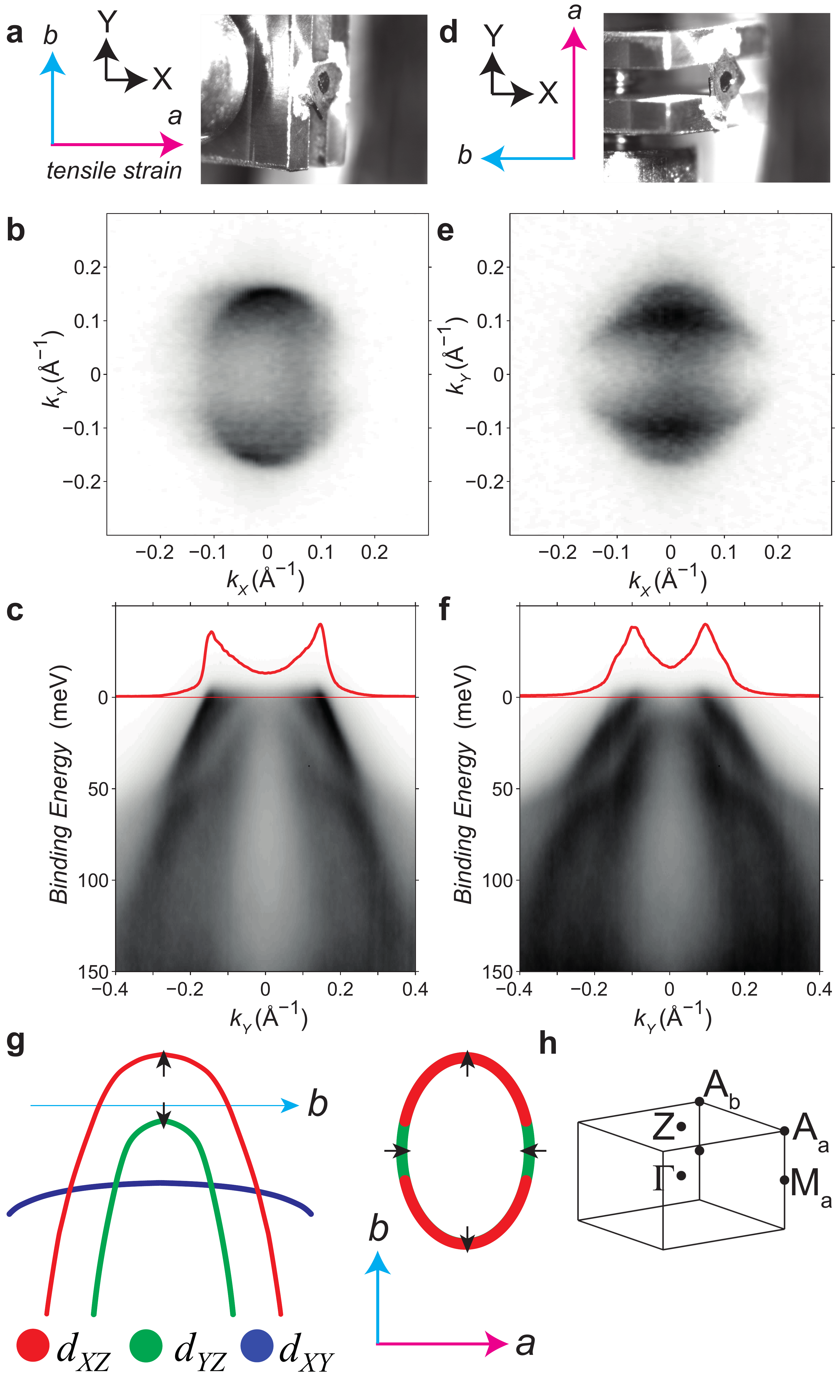}
	\caption[Hole Pockets]{\textbf{ARPES measurements of the hole pockets of detwinned FeSe}. a) Microscope image of the sample in-situ after cleavage, with tensile strain provided by the horseshoe device. b) Fermi surface map around the Z point, and c) high-symmetry cut in the $\rm Z-A_b$ direction, also plotting the MDC at the Fermi level. d-f) As left, but with sample rotated by 90$^\circ$ with respect to the measurement geometry, such that the high-symmetry cut is now in the $\rm Z-A_a$ orientation. All data obtained at 23 eV in LH polarisation, which highlights the $d_{XZ}$ orbital weight here. g) Simplified sketch of the hole band dispersions and Fermi surface within one domain. Arrows indicate the shifts observed in the nematic phase compared with the high-temperature tetragonal phase. h) Labelled Brillouin zone of FeSe.}
	\label{fig1}
\end{figure}

\textit{Hole Pockets.} In Fig.~\ref{fig1} we present detwinned ARPES spectra of the hole pocket of FeSe, using a photon energy of 23 eV which corresponds to the Z point at the top of the Brillouin zone \cite{Watson2015a}, where the hole pocket is largest and the detwinning effect can be best resolved. The ARPES data are obtained with equivalent measurement geometries, with only the azimuthal orientation of the sample differing by 90$^\circ$. Throughout this paper we plot only the photoemission intensity, and do not rely on any second-derivative analysis. To avoid ambiguity we use $X,Y$ labels when referring to the measurement geometry, while $x,y$ are defined with respect to the orthorhombic $a,b$ axes. We determine the $a$ and $b$ axes by associating the longer $a$ axis with the direction of the tensile uniaxial strain. It has already been deduced from measurements of twinned samples that there is a single elliptical hole-like Fermi surface at low temperatures \cite{Watson2015a,Watson2016,Fedorov2016}. However only detwinned measurements can determine the direction of the elongation of the hole pocket with respect to the orthogonal axes: by comparing the Fermi surface maps in Fig.~\ref{fig1}b,e), we find that the longer axis of the elliptical hole pocket is directed along the shorter $b$ crystallographic axis in the orthorhombic phase (consistent with the result of Ref.~\cite{Suzuki2015}). This implies that at the $\Gamma$ (or Z) point in the nematic phase, the $d_{xz}$ orbital is raised in energy, whereas the $d_{yz}$ orbital is lowered \cite{Watson2017c}. Since it is well-established that the electron band with $d_{yz}$ character is raised up towards the Fermi level at the M point \cite{Shimojima2014,Suzuki2015,Watson2015a,Watson2016}, this results confirms that a momentum-independent ordering of $d_{xz/yz}$ orbitals (i.e. ferro-orbital ordering) is not a possible explanation of the data \cite{Suzuki2015,Watson2016}, and constrains any other theoretical description of nematic order.  

\begin{figure*}
	\centering
	\includegraphics[width=\linewidth]{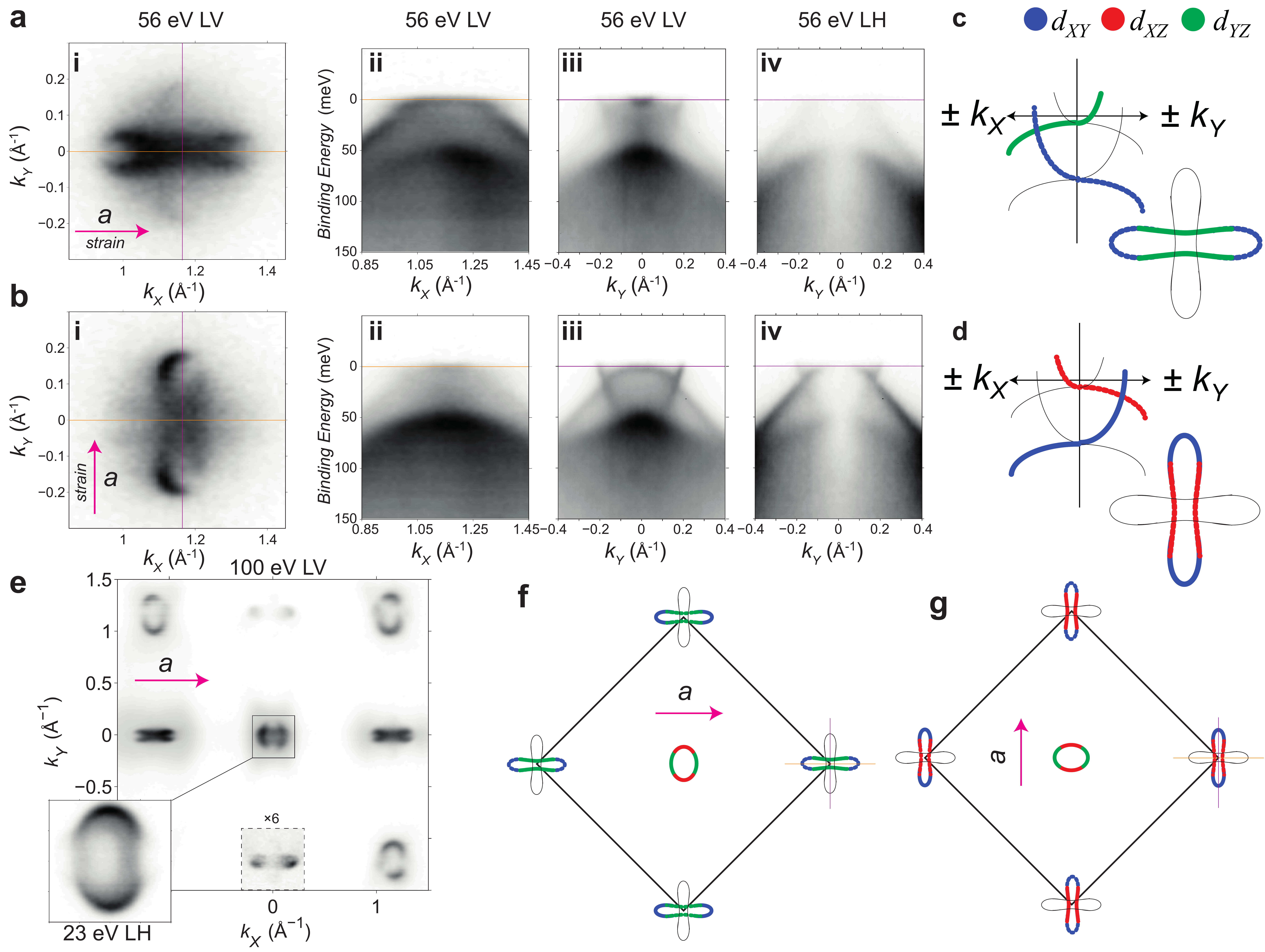}
	\caption[Electron Pockets]{\textbf{ARPES measurements of the electron pockets in one domain.} a) i) Fermi surface map around the A point. ii-iii) High symmetry cuts along $k_X$ and $k_Y$ obtained with 56 eV LV (vertical) polarisation, and iv) along $k_Y$ with LH (horizontal). b) Equivalent measurements to above, but with the sample rotated by 90$^\circ$. c,d) Schematic of the observed bands in each domain; dashed lines represent bands observed with 'switched parity' and thin black lines represent the weak intensity contributions from the minority domain. e) Fermi surface map on an `accidentally detwinned sample' taken at 100 eV in LV polarisation. Note that although the selection rules alternate between the first and second $\bar{\Gamma}$ points, the elongation of the hole pocket is along the $b$ direction in both locations. This is clarified in the inset which shows a detailed map of the hole Fermi surface at 23 eV LH on this sample which compares with Fig.~\ref{fig1}b), indicating a detwinning effect in this sample. f,g) Schematic apparent Fermi surfaces observed by ARPES measurements for the two orientations.}
	\label{fig2}
\end{figure*}

It can be seen from the high-symmetry measurements in Fig.~\ref{fig1}c,f) that although the intensity primarily follows either the inner or outer dispersion depending on the sample orientation, a much weaker intensity remains on the minority domain. This indicates that the sample is not fully detwinned, but rather that the twin population is heavily weighted towards one orientation due to the applied uniaxial strain. Furthermore we can quantitatively estimate the detwinning effect by comparing the amplitude of peaks derived from fitting the MDCs shown in Fig.~\ref{fig1}c,f). By fitting the Momentum Distribution Curves (MDCs) at the Fermi level with the same peak widths in both cases, the amplitudes are related to the domain populations via a simple relationship described in the SI, and from this we estimate that the domain populations in this sample are split 80\%-20\%.

\textit{Electron Pockets.} In the high temperature tetragonal phase, the Fermi surface of the electron pockets around the M or A points consists of two crossed ellipses \cite{Watson2016}. These ellipses undergo distortions in the nematic phase, and a cross-shaped feature consisting of two peanut-shaped electron pockets is observed at low temperatures in ARPES measurements of twinned samples, which has been interpreted within various different schemes \cite{Shimojima2014,Watson2015a,Watson2016,Suzuki2015,Fedorov2016,Fanfarillo2016}. However until now it has not been clearly determined how the electron bands would appear within one domain. In Fig.~\ref{fig2}a-b) we find a remarkable result: \textit{the ARPES measurements only show intensity on the one peanut-shaped Fermi surface which is oriented along the longer $a$ axis}. This can be considered as a manifestation of nematic order: the system has chosen to pick out a unique direction, and fourfold symmetry of the spectral intensity is completely lost.  

In Fig.~\ref{fig2}c,d) we show a schematic of the observed bands deduced from the maps and cuts presented in each sample orientation, corresponding to a single peanut-shaped Fermi surface in each case. The orbital characters are based on the clear identification of separate $d_{XY}$ and $d_{XZ/YZ}$ bands in Ref.~\cite{Watson2016}. The contribution of the minority twin domain to the observations is drawn as a thin black line. We do not draw the expected second electron pocket along the $b$ axis, which is not directly observed by ARPES at low temperature.
 
This extraordinary result is accompanied by other stringent and peculiar selection rules which modulate the observed spectral weight. In FeSe the bands have distinct orbital characters, and in the simplest case, by considering the parity of the relevant atomic $d$-orbital with respect to the scattering plane, one can infer whether or not any photoemission intensity is to be expected for a given incident beam polarization \cite{Wang2012}. According to these basic selection rules we would expect the bands with $d_{XZ}$ and $d_{YZ}$ orbital characters to be observed in ``LH" ($p$) and ``LV" ($s$) polarizations respectively, which in fact holds true at normal emission \cite{Watson2015a}. However due to the presence of a glide symmetry connecting the two Fe sites in the unit cell, some bands pick up an additional phase factor in experiments \cite{Lin2011,Brouet2012,Tomic2014,Moreschini2014}, and are then observed with the opposite polarisation to the conventional expectation. This is highly relevant for the understanding of the observations of the electron pockets in detwinned samples, although the effect is not specific to the nematic phase. For instance, the vertical sides of the Fermi surface in Fig.~\ref{fig2}b) have $d_{XZ}$ character, but they are observed with ``switched parity" and seen in LV polarization only, where conventionally the $d_{XZ}$ orbital should be completely suppressed. In Fig.~\ref{fig2}e) we show a large area Fermi surface map obtained from a different sample, which was not mounted on a strain device, but appears to have been ``accidentally detwinned", perhaps due to an anisotropic strain from the gluing of the sample \footnote{Out of more than 30 samples measured this is the only example where a significant accidental detwinning was observed. An equivalent Fermi surface measurement of a twinned sample is presented in the Supplementary Information and shows clear qualitative differences in terms of which bands are observed.}. The $\rm M_a$ point (i.e. the M point reached along the $a$ axis) on the scattering plane is similar to Fig.~\ref{fig2}a-i), but at the $\rm M_b$ point the intensity is reversed - the strong $d_{YZ}$ intensity is completely suppressed, but the ends of the pocket with $d_{XY}$ character are observed. Moreover at the hole pockets in the second Brillouin zone, the $d_{XZ}$ sections appear in LV polarization while the $d_{YZ}$ sections are suppressed, the opposite case to the first $\Gamma$ point where it is the $d_{YZ}$ sections which are observed conventionally. We can summarise the matrix element effects by observing that the simple, conventional matrix elements arguments based on the parity of the d-orbital apply at normal emission, but that as a result of interference effects arising from the glide symmetry which relates neighbouring Fe sites, any translation of the final photoelectron momentum by the wavevector ($\pi$,$\pi$) induces a “parity switching”, such that a band with a given orbital character is observed with the opposite polarisation to the conventional expectation. This parity switching behavior persists even into the 3\textsuperscript{rd} and 4\textsuperscript{th} Brillouin zones (SI). Full schematics of the Fermi surface of FeSe as measured by ARPES are represented in Fig.~\ref{fig2}f,g).

\begin{figure}
	\centering
	\includegraphics[width=0.8\linewidth]{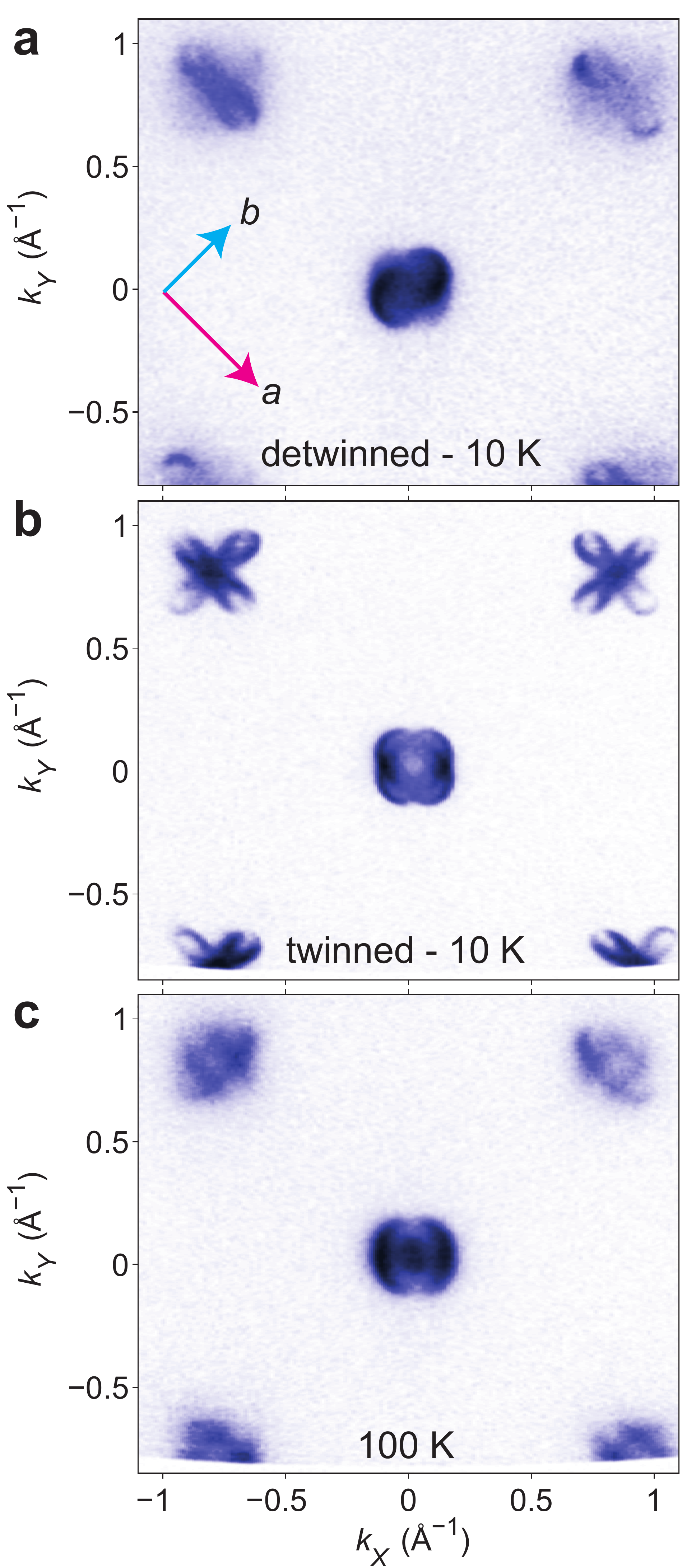}
	\caption[Big Maps]{\textbf{Fermi surface maps of FeSe.} Data are taken at 56 eV, LV polarisation, approximately corresponding to a Z-A plane. a) Fermi surface map of detwinned FeSe. b) Equivalent measurement of an unstrained twinned sample, including the characteristic cross-shaped intensity at the M point seen in most measurements. c) Measurement at 100~K in the tetragonal phase in the unstrained sample, showing the full structure of the electron pockets.}
	\label{fig3}
\end{figure}

 In Fig.~\ref{fig3} we summarise the comparison between measurements of twinned and detwinned samples with Fermi surface maps covering the whole first Brillouin zone. For the detwinned sample shown in Fig.~\ref{fig3}a), the electron pockets at each A point display intensity only on the peanut oriented along the $a$ axis. Moreover, the parity switching behaviour is also seen in the different structures observed for the electron pockets at A points separated by ($\pi$,$\pi$). These measurements in an unbiased 45$^\circ$ rotated geometry confirm that the selection rules obeyed by the electron pockets are intrinsically related to the sample orientation as drawn in Fig.~\ref{fig2}f,g), and are not a special case due to the high symmetry measurement geometry. It is natural that the equivalent measurements of twinned samples in Fig.~\ref{fig3}b) can be understood as a superposition of contributions from both domains. However what is less intuitive is that one must also consider this superposition of matrix element effects to account for the data at 100 K in Fig.~\ref{fig3}c) in the tetragonal phase. As a result of this superposition, essentially the whole Fermi surface may be observed at high temperatures, however within one orthorhombic domain this is not the case.  

\section{DISCUSSION} 

The observed parity switching amongst the bands with $d_{xz}/_{yz}$ orbital character which accompanies any translation by ($\pi$,$\pi$) is due to the fact that these orbitals have odd parity under the glide symmetry which relates neighboring Fe sites \cite{Brouet2012}. This effect is therefore not related to the nematic order, and is also observed in the tetragonal phase. However the orientation of the orthorhombic distortion seems to play a crucial role in the ARPES selection rules, since the longer $a$ axis picks out the direction of the branch of the electron pocket on which spectral weight is found, whereas all bands are observed in the tetragonal phase. The $Cmma$ unit cell suggested for orthorhombic FeSe \cite{Khasanov2010} is only weakly distorted compared to the $P4/nmm$ tetragonal phase and preserves most of the symmetry elements, so such a dramatic change in the observations is unexpected. We suggest that there must exist a much deeper source of anisotropy in the nematic phase, which then manifests in the ARPES selection rules. Therefore in one sense, ARPES measurements reveal an electronic Ising-nematic order parameter via the matrix elements, as well as determining the symmetry-breaking quasiparticle band shifts in the nematic phase \cite{Watson2016}. The link between this ``one-peanut" selection rule and the underlying nematic order is a fascinating question which remains to be solved, but speculatively this could be related to a fluctuating  antiferromagnetic order.

The fate of the electron pocket oriented along the $b$ axis in the nematic phase remains something of a puzzle. One possible interpretation could be that the non-observation of this pocket may be a strong selection rule, specific to the technique of ARPES. In this case the pocket still exists, still contributes to the Luttinger count, and may have a detectable influence on e.g. the magnetotransport properties. However intriguingly, recent quasiparticle interference (QPI) \cite{Sprau2017Science} measurements also only detected a single peanut-shaped electron pocket in FeSe in the low-temperature limit. In that case the non-observation of the peanut along $b$ was attributed to a dramatically lower quasiparticle weight on the $d_{xz}$ and $d_{xy}$ orbitals \cite{Kreisel2017} which predominantly constitute that pocket, while the $d_{yz}$ orbital which dominates the electron pocket along $a$ was proposed to remain coherent with near-unity quasiparticle weight. Broadly speaking, this scenario could be considered to be consistent with our ARPES measurements, but some details vary. For instance, we can clearly observe some $d_{xy}$ dispersions on the electron pocket that we do observe in Fig.~\ref{fig2}(b). Moreover we find sharp quasiparticle-like dispersions of the $d_{xz}$ sections of the hole pocket in Fig.~\ref{fig1}(c), and we do not see any obvious evidence for an orbitally-selective loss of coherence around the hole pocket. Nevertheless a pocket- or momentum-selective loss of quasiparticle weight would be a reasonable alternative explanation of our measurements. However given the high resolution of the measurements here and in the literature, it would be necessary for the quasiparticle weight to be extremely low all around this pocket (e.g. less than the observed weak intensity from the $\sim$20$\%$ minority twin domain), otherwise it would have been directly detected in some geometry.  Firm arguments for either scenario are yet to be developed and other explanations could be possible, and so we would encourage further theoretical investigations of this phenomenon.   

The knowledge gained by these detwinned ARPES measurements complements the existing high-resolution ARPES measurements of twinned samples of FeSe \cite{Watson2017c,Watson2016}, and allows for some robust conclusions on the the experimental electronic structure of FeSe in the nematic phase, in conjunction with quantum oscillations \cite{Terashima2014,Watson2015b} and QPI \cite{Sprau2017Science} data in the literature. Here we have confirmed that the single hole and the single observed electron pocket are oriented orthogonal to one another in the nematic phase, which cannot be explained with a simple $n_{xz}-n_{yz}$ ferro-orbital polarisation on the Fe sites, and is instead a firm proof that momentum-dependent or bond-type orderings must be considered. Thus a purely ferro-orbital ordering is excluded, and moreover a purely $d$-bond nematic ordering is excluded because such a scenario cannot account for the band splittings at the $\Gamma$ point \cite{Watson2017c,Coldea2017Review_arxiv}. The unidirectional nematic bond ordering which we proposed in Ref.~\cite{Watson2016} naturally accounts for the observed experimental Fermi surfaces, but it is a now a rather subtle question whether differences between this and other proposals which vary mainly in the shape and size of the non-observed electron pocket \cite{Kreisel2017,Onari2016} can be distinguished experimentally. However we emphasise that the dramatic one-peanut observation revealed by our measurements presents a more fundamental question than the exact details of the magnitude and momentum-dependence of the orbital ordering in the nematic phase. 

To summarise, we have reported high-resolution detwinned ARPES measurements of FeSe. At low temperatures, the observed Fermi surface consists of a small elliptical hole pocket oriented along the $b$ axis and a small peanut-shaped electron pocket oriented along the $a$ axis; the expected electron pocket along the $b$ axis is not observed at all. Our result mark a step-change in the quality of ARPES spectra of detwinned Fe-based superconductors and reveal the profound electronic anisotropies in the nematic phase of FeSe.

\section{Acknowledgments}

The authors are grateful to A.~I.~Coldea for initial discussions which led to this work. We thank S.~V.~Borisenko, V.~Brouet, M.~Eschrig, R.~M.~Fernandes and R.~Valenti for useful discussions. We thank Diamond Light Source for access to beamline I05 (proposal numbers CM12153,NT15663) that contributed to the results presented here. A.A.H. acknowledges the financial support of the Oxford Quantum Materials Platform Grant (EP/M020517/1). L. C. R. is supported by an iCASE studentship of the UK Engineering and Physical Sciences Research Council (EPSRC) and Diamond Light Source Ltd CASE award.	

\clearpage

%

\clearpage
\end{document}